\documentstyle[floats,prl,aps]{revtex}

\begin{document}

\title{Exact Large Deviation Function in the Asymmetric Exclusion Process}

\author{Bernard Derrida$^{\dag }$  and
Joel L. Lebowitz$^{\ddag}$}
\address{$\dag$ Laboratoire de Physique Statistique,
 24 rue Lhomond, 75231 Paris Cedex 05,
 France; {\rm email derrida@physique.ens.fr}\\
$\ddag$  Department of Mathematics,
Rutgers University, New Brunswick, NJ 08903; 
 {\rm email lebowitz@math.rutgers.edu}}

\author{\parbox{397pt}{\vglue 0.3cm \small
By an extension of the Bethe ansatz method used by Gwa and Spohn, we
obtain an exact expression for  the large deviation function of the time
averaged current for the fully asymmetric exclusion process in a ring 
containing $N$ sites and $p$ particles.  Using this expression we
easily recover the exact diffusion constant obtained earlier
and calculate as well some higher cumulants.  The  distribution of
the deviation  $y$ of the average current
is, in the limit $N \to \infty$, skew and decays like
$\exp -  (A y^{5/2})$ for $y \to + \infty$ and $\exp - (A' |y|^{3/2})$ for
$y \to -\infty$. 
Surprisingly,  the large 
deviation function has an expression very similar to  the 
pressure (as a function of the density) of an ideal Bose or Fermi gas
in $3d$.\
}}
\maketitle

\bigskip

In the last decade, it has become clear that questions as different as
$1d$ turbulence, growth of interfaces or directed polymers in a random
medium in dimension $1+1$ dimension were different versions of the
same problem described by the noisy Burgers or equivalently the
Kardar-Parisi-Zhang (KPZ) equation \cite{HHZ,bmp}.  One of the
simplest lattice versions of this problem is the asymmetric exclusion
process (ASEP) which also has a long history both in the mathematical
and in the physical literature \cite{Lig,Spo,Krug}.  It describes a
driven lattice gas with hard core exclusion and under suitable
scalings the evolution of its macroscopic density is governed by a
noisy Burgers equation in $1d$ which is equivalent to the KPZ
equation.

Many properties of the ASEP have been calculated (steady states, mass
gaps, diffusion constants) by various methods including exact
solutions \cite{mrsb,dhar,gwa,DEHP,dem,ss,gs,janowsky} and compared
\cite{HHZ,kt} with the predictions of the replica approach
\cite{kardar,zhang,bo} or numerical simulations \cite{HH,bray,kmhh}
for directed polymers or growth problems.

Here, extending the Bethe ansatz approach of Gwa and Spohn \cite{gwa},
we obtain the expression for a large deviation function which gives
the exact probability for the time averaged current to take any,
typical or untypical, value.  Similar distributions have been studied
numerically in the past \cite{HH,bray,kmhh}, in particular to better
understand the $n \to 0$ limit in the replica approach of directed
polymers in a random medium.  Other recent results include the
numerical calculation of the distribution of the density for a
randomly driven Burgers equation \cite{yc}.  The results presented
below are, to our knowledge, the first exact determination of a
non-trivial probability distribution for a fluctuating quantity
described by the noisy Burgers equation or equivalently the KPZ
equation.  Through the known equivalences between the different
problems mentioned above, our results could easily be used to obtain
the whole distribution of the ground state energy of a directed
polymer in a random medium in dimension 1+1 for a finite lateral
geometry.

We consider here a totally ASEP consisting of $p$ particles on a ring
of $N$ sites.  During any time interval $dt$, each particle jumps with
probability $dt$ to its right if the target site is empty.  The unique
stationary state of this system is one in which all ${N\choose p}$
different configurations $C$ have equal weight \cite{mrsb}.  We call
$Y_t$ the {\it total} distance covered by {\it all the particles }
between time $0$ and time $t$ (thus $Y_0 = 0$).  The probability
$P_t(C)$ of finding the system in configuration $C$ at time $t$
satisfies
$$ {d \over dt}P_t(C) = 
 \sum_{C'} [M_0(C,C')
 +  M_1(C,C')] P_t(C') $$
where $M_1(C,C') dt$ is the probability of going from $C'$ to $C$
during a time interval $dt$, increasing $Y$ by 1, while $M_0$ is a
diagonal matrix,
$$M_0(C,C) = - \sum_{C' \neq C} M_1(C',C) $$

Let $P_t(C,Y)$ be the joint probability of being, at time $t$, in the
configuration $C$ and having $Y_t = Y$, some non-negative integer.  Define
also $F_t(C)$ as $\sum_{Y=0}^\infty \exp(\alpha Y) P_t(C,Y)$.  Then
$$
{\partial F_t(C) \over \partial t} = \sum[M_0(C,C') + e^\alpha M_1(C,C')]
F_t(C')
$$
Summing $F_t(C)$ over $C$ we obtain $\langle e^{\alpha Y_t} \rangle$.  It
is easy to see that 
for large $t$
\begin{equation}
\langle e^{\alpha Y_t} \rangle \sim  e^{  \lambda(\alpha)  t}
\label{exp}
\end{equation}
where $\lambda(\alpha)$ is the largest eigenvalue of the matrix
$M_0 + e^\alpha M_1$  (we have omitted the dependence of $\lambda$ on
$p$ and $N$).  By $\sim$ we mean:  take the logarithm of
both sides of (1), divide by $t$ and let $t \to \infty$ to obtain an
equality.  While (1) holds  for every initial configuration, it is
sometimes convenient to think of the initial state $P_0(C)$ as the
stationary one. 

In the present paper we show that the exact expression of 
 $\lambda(\alpha)$, 
 for  arbitrary $p$
and $N$, is given by the power series in $\alpha$ obtained by
eliminating $B$ between the  following two series
\begin{equation}
\lambda(\alpha)= - p \sum_{q=1}^\infty B^{q}  {(Nq-2)! \over (pq)! \
(Nq-pq-1)!}  
\label{sol1}
\end{equation}
\begin{equation}
 \alpha = -   \sum_{q=1}^\infty B^{q}  {(Nq-1) ! \over (pq)! \  (Nq-pq)!}  
\label{sol2}
\end{equation}

These series have a non-zero radius of convergence in $B$ so
$\lambda(\alpha)$ is analytic in $\alpha$ at least near
$\alpha=0$. Moreover, by the Perron-Frobenius theorem there is no crossing
of the largest eigenvalue of the matrix $M_0 + e^\alpha M_1$ for real
$\alpha$ and fixed finite $N$.  Hence (\ref{sol1}) and (\ref{sol2})
determine $\lambda(\alpha)$ on the 
whole real axis as the analytic continuation of the expression in
the vicinity of $\alpha=0$.  In the rest of this work, we first show how
(\ref{sol1}, \ref{sol2}) can be obtained using the Bethe ansatz. Then we
investigate some 
consequences of this exact expression.

The Bethe ansatz we use   is very similar to that used by 
 Gwa and Spohn  \cite{gwa}, except that here there is an extra
parameter $\alpha$. 
Specifying a configuration $C$ by the positions 
  $1 \leq x_1< x_2 < ...< x_p \leq N$  of the $p$ particles we write an
eigenvector of the matrix $M_0 + e^\alpha M_1$ in the form 
\begin{equation}
\sum_Q A_Q \prod_{j=1}^p [z_{Q(j)}]^{x_j} 
\label{eigenvector}
\end{equation}
where the sum is over all the permutations $Q$ of $1,2,...,p$ and the $p$ 
parameters $z_k$ satisfy the following equations 
\begin{equation}
 {( e^\alpha-z_k)^p \over z_k^N} = (-)^{p-1}  \prod_{j=1}^p (e^\alpha-z_j) 
\label{eqz1}
\end{equation}
\begin{equation}
 z_1 z_2 ... z_p =1
\label{eqz2}
\end{equation}
The derivation of (\ref{eqz1},\ref{eqz2}) is almost identical to that
given by what
Gwa and Spohn in \cite{gwa} and we refer to their work.

For any solution $\left\{ z_k \right\}$ of (\ref{eqz1},\ref{eqz2}), expression 
(\ref{eigenvector}) gives an eigenvector of the matrix $M_0+ e^\alpha
M_1$ with eigenvalue $\lambda(\alpha)$ given by 
\begin{equation}
\lambda(\alpha) = 
e^\alpha \left[ {1 \over z_1} + {1 \over z_2} + ... + {1 \over z_p}
\right] - p 
\label{lambda}
\end{equation}

The difficulty is that
there are  in general many solutions of (\ref{eqz1},\ref{eqz2})
 corresponding to different eigenvalues of the matrix $M_0 +
 e^\alpha M_1$. The one we need to choose is the 
largest eigenvalue $\lambda(\alpha)$ which, as already noted,  can be
followed when $\alpha $ 
varies,
from  the unique eigenvalue which vanishes  when $\alpha \to 0 $. 
Thus $\lambda(\alpha)$ corresponds
 to the solution such that all the $z_k \to 1$ as $\alpha \to 0$.

To see why (\ref{sol1},\ref{sol2}) are equivalent to
(\ref{eqz1},\ref{eqz2},\ref{lambda}) when one chooses the above solution,
we consider a function 
$h(z)$ analytic in the circle $|z-e^\alpha| < 2
\epsilon$.  It is then  an easy consequence of the Cauchy theorem that
\begin{equation}
\sum_{1 \leq k \leq p} h(Z_k) = \oint {dz \over 2 \pi i z}    h(z)
{(z- e^\alpha)^{p-1}  [ p z -N ( z- e^\alpha )] 
\over
(z- e^\alpha)^p   -A z^N} 
\label{identity}
\end{equation}
where the contour is the circle $|z - e^\alpha| = \epsilon$, the
parameter $A$ is such that $|A| \ll \epsilon^p e^{-N\alpha}$ and the
$Z_k$ are the  $p$ solutions of 
\begin{equation}
(z-e^\alpha)^p  = A z^N
\label{eqZ}
\end{equation}
 which tend to $e^\alpha$ as $A \to 0$.
As (\ref{eqZ}) has a form identical to (\ref{eqz1}), one can use
(\ref{identity}) 
to calculate $\lambda(\alpha)$ (by choosing $h(z)= -1 + e^\alpha /z$)
and this gives 
(\ref{sol2}) in terms of $B = A e^{(N-p)\alpha}$. The constant $B$ can
then be determined by 
requiring that (\ref{eqz2}) is also satisfied. To do that one can use again
the identity (\ref{identity}) for the function $h(z)=\log z$ and this
gives (\ref{sol2}).

{}From the exact expressions  (\ref{sol1},\ref{sol2}),
one can expand $\lambda(\alpha)$ in a power series in $\alpha$.
In particular,  one can check   that 
\begin{eqnarray}
 \lambda(\alpha) =  {p(N-p) \over N-1}  \alpha +   \ \ \ \ \ \ \ \ \ \ 
\ \ \ \ \ \ \  \ \ \ \ \ \ \ \ \  \ \ \ \ \ \ \ \ \ \nonumber \\
\ \ \    { p \  (p!)^2   \  (N-p)!^2 \ (2N-3)! \over (2p!) \  (N-1)!^2
\ (2N-2p-1)!  } \alpha^2 +O(\alpha^3)  
\label{dev}
\end{eqnarray}
and using (\ref{exp}), this gives
\begin{eqnarray}
\lim_{t \to \infty} {\langle Y_t \rangle \over t} = \left. {d
\lambda(\alpha) \over d \alpha } \right|_{\alpha=0} = { p (N-p) \over
N-1}      
\label{cu1} 
\end{eqnarray}
which is known to be  a simple consequence of the fact that in  the
steady state 
all configurations have equal weight \cite{mrsb,dhar}.
(Note that if $P_0(C)$ was the stationary measure,
(\ref{cu1}) would in fact  be valid for all $t$.) 
The expansion (\ref{dev}) also gives the diffusion constant
\begin{eqnarray}
  \lim_{t \to \infty} { \langle Y_t^2 \rangle
 - \langle Y_t \rangle^2
 \over t} = \left. {d^2 \lambda(\alpha) 
\over d \alpha^2 } \right|_{\alpha=0} 
\label{cu2} 
\end{eqnarray}
which agrees  (up to a trivial factor $p^2$ due to a slight change in
the definition of $Y_t$) with 
the  expression  
 obtained by an extension \cite{dem} of the matrix method \cite{DEHP}.
Exact expressions of higher cumulants of $Y_t$ may be obtained in a
similar way from (\ref{exp},\ref{sol1},\ref{sol2}) by expanding
$\lambda(\alpha)$ to higher orders in $\alpha$. 
 For example, the third derivative of $\lambda$ at $\alpha=0$ gives
 \begin{eqnarray}
 \lim_{t \to \infty}{\langle Y^3 \rangle - 3\langle Y \rangle^2
\langle Y \rangle 
 + 2 \langle Y \rangle^3 \over t}   = \ \ \ \ \ \ \ \ \ \ \ \ \ \ \ \
\ \  \ \ 
 \label{cu3} 
 \\
 \ \   {3 p (N-p)^3  \over  (2N-1) (N-1)} {2N-1 \choose 2p}^2 {N-1
\choose p}^{-4}  \nonumber \\ 
    \ \ \ \ \ \ \ \ \ \ - {4 p (N-p)^3\over  (3N-1)(N-1)} {3N-1
\choose 3p} {N-1 \choose p}^{-3}  
 \nonumber
 \end{eqnarray}

For  large positive  $\alpha$ (i.e. $\alpha  \gg 1$), the solutions
$z_k$ of (\ref{eqz1},\ref{eqz2}), for $1\leq k \leq p $  are given by 
$$z_k = e^{(2k-p-1)\pi i /N}
+O(e^{-\alpha})
$$  
and this  leads to
\begin{equation}
\lambda(\alpha) = e^\alpha \; {\sin (p \pi /N) \over \sin(\pi / N)}
+O(1)
\label{largeap}
\end{equation}
For  large negative $\alpha$  (i.e.  $-\alpha  \gg 1$), one of the
$z_k$ becomes  large 
(say $z_1\simeq e^{(1-p)\alpha}$) and  the $p-1$  other $z_k$ become  small
($z_k \simeq e^\alpha$ for $2 \leq k \leq p$). This leads to
\begin{eqnarray}
\lambda(\alpha) \simeq  \left\{
\begin{array}{c}
  -1 + e^{p\alpha}  \ \ \ \makebox{for } \ \ p < N-p  \\
  -1 + 2 e^{p\alpha} \ \ \  \makebox{for } \ \ p = N-p  \\
  -1 + e^{(N-p) \alpha}  \ \ \  \makebox{for } \ \ p > N-p 
\end{array}
\right.
\label{largean}
\end{eqnarray}
(We can understand the limit $\alpha \to -\infty$ in (\ref{largean}) by
noting that the probability of not making any jump during the time
interval $t$ is, for a configuration in which all particles form one
cluster, just $\exp(-t)$.)

In the limit of large  $N$ and $p$, keeping the ratio
$p/N= \rho$ fixed, one finds for the  first four cumulants of $Y$ 
(by calculating the first four derivatives of $\lambda(\alpha)$ at $\alpha=0$)
\begin{equation}
\lim_{t \to \infty}{\langle Y \rangle \over t}  \simeq 
 N \rho(1-\rho)  
\label{mom1}
\end{equation}
\begin{equation}
\lim_{t \to \infty}{\langle Y^2 \rangle - \langle Y \rangle^2 \over t}   \simeq
 N^{3/2}  [ \rho (1-\rho)]^{3/2}  {\sqrt{\pi} \over 2}  
\label{mom2}
\end{equation}
\begin{eqnarray}
\lim_{t \to \infty}{\langle Y^3 \rangle - 3\langle Y^2 \rangle \langle
Y \rangle 
+ 2 \langle Y \rangle^3 \over t}  \ \ \ \ \ \ 
\nonumber \\  \ \ \ \ \ \ \  \simeq
 N^3 [\rho (1-\rho)]^{2}    \left( {3 \over 2} - {8 \over 3^{3/2}} 
\right)  \pi
\label{mom3}
\end{eqnarray}
\begin{eqnarray}
\lim_{t \to \infty}{\langle Y^4 \rangle - 3\langle Y^2 
\rangle^2 -4 \langle Y^3 \rangle
\langle Y \rangle +12 \langle Y^2 \rangle \langle Y \rangle^2
- 6 \langle Y \rangle^4 \over t} 
\nonumber \\ \simeq
 N^{9/2} [\rho (1-\rho)]^{5/2} 
 \left( { 15 \over 2} + {9 \over   2^{1/2}} - {24\over 3^{1/2}}\right) 
  \pi^{3/2} 
\label{mom4}
\end{eqnarray}

In fact, in this limit $N \to \infty$ with $\rho = p/N$ fixed,
the expressions
(\ref{sol1}) and (\ref{sol2}) can be rewritten as 
\begin{eqnarray}
\sqrt {2\pi N^3 \over \rho(1-\rho)} \left [\lambda(\alpha) - \alpha\  {p
(N-p) \over N} \right ] \simeq  \ \ \ \ \ \ \  \nonumber \\
 \ \ \ \ \ \ \ G(\beta) \  \equiv 
\ \sum_{q \geq 1} { (-)^{q+1} C^q \over q^{5/2}} 
\label{sol1a}
\end{eqnarray}
where
\begin{equation}
\alpha \; \sqrt{2\pi \rho(1-\rho)N^3}   \    \simeq \   \beta \  \equiv \ 
\sum_{q \geq 1} { (-)^{q+1} C^q \over q^{3/2}} 
\label{sol2a}
\end{equation}
and  $C= - B N^N p^{-p} (N-p)^{-N+p}$.
This shows that for $N$ large and $\beta$ fixed,
 i.e. for $\alpha \sim N^{-3/2}$,
the difference $\lambda(\alpha) - \alpha p (N-p)/N$ takes  a scaling form
\begin{equation}
\lambda(\alpha) - \alpha\  {p (N-p) \over N} \simeq  
\sqrt{\rho (1- \rho)  \over 2 \pi N^3 } 
G \left( \alpha  \sqrt{2 \pi \rho (1- \rho) N^3} \right) 
\label{scaling}
\end{equation} 

The two sums which define $G(\beta)$  in the right hand side of
(\ref{sol1a},\ref{sol2a})  
are very reminiscent of the pressure and of the
density of an Ideal Bose or Fermi gas
(depending on the sign of $C$)
in 3 dimension \cite{huang}: positive $C$ and $\alpha$ correspond to the 
Fermi gas and negative $C$ and $\alpha$ correspond to the Bose gas.

The range $ N^{-3/2} \ll \alpha \ll 1$ (i.e. large positive $C$)
correponds to the high pressure or low  
temperature limit of a Fermi gas and  this analogy  leads to
\begin{equation}
\lambda(\alpha) - \alpha \ {p (N-p) \over N} \simeq {(3 \pi)^{2/3} \over 5}
[ \rho(1-\rho)]^{4/3} N \alpha^{5/3}   
\label{ap}
\end{equation}
Negative $\alpha$ can also be analysed:
when $C$ varies from $0$ to $-1$ in
(\ref{sol1a},\ref{sol2a}), $\alpha$ decreases
(this corresponds to a Bose gas in its high temperature phase) 
and $\alpha$ has a finite limit $\alpha_0$ as $C\to -1$.
To decrease further $\alpha$, it turns out that
 one has to increase $C$ from $-1$ to $0$ and
to replace  the expressions (\ref{sol1a}) and (\ref{sol2a})  of
$G(\beta)$ and $\beta$ by 
their analytic continuations \cite{dz}
\begin{equation}
G(\beta)  =  
{4 \over 3}  \sqrt{\pi} [ - \log (- C)]^{3/2} +
 \ \sum_{q \geq 1} { (-)^{q+1} C^q \over q^{5/2}}  
\label{sol1b}
\end{equation}
\begin{equation}
\beta =
- 4 \sqrt{\pi} [ - \log (- C)]^{1/2} + 
\ \sum_{q \geq 1} { (-)^{q+1} C^q \over q^{5/2}}  
\label{sol2b}
\end{equation}
This analytic continuation is a major difference with the usual
Bose-Einstein condensation
where $C$ remains fixed at the value $-1$ in the whole low temperature phase. 
Here, for $\alpha < \alpha_0$, one of the $z_k$ becomes much larger
than the others but, in contrast to the Bose condensate,
the chemical potential $C$ in (\ref{sol1b}) keeps varying 
with $\alpha$.  For the range $ N^{-3/2} \ll  - \alpha \ll 1$, this leads to
\begin{equation}
\lambda(\alpha) - \alpha \ {p (N-p) \over N} \simeq {-1 \over 24} [
\rho(1-\rho)]^{2} N^3 \alpha^{3}    
\label{an}
\end{equation}

The knowlegde of the function $\lambda(\alpha)$ determines
(through $y=d \lambda/ d \alpha -p(N-p)/N$ and
 $f(y)=\lambda -  \alpha [y+p(N-p)/N]$) 
the large deviation function
$f(y)$ defined here as
\begin{equation}
f(y) = \lim_{t \to \infty} {1 \over t} \log \left\{ \makebox{Prob} \left(
{Y_t \over t} - N \rho (1-\rho) = y \right) \right\} 
\label{devia}
\end{equation}
{}From (\ref{mom1},\ref{mom2},\ref{scaling},\ref{ap},\ref{an}) one can
show that 
for $N$ large, $f(y)$ takes the  scaling form for $|y| \ll N$
\begin{equation}
f(y) \simeq \sqrt{\rho(1-\rho) \over \pi N^3}
 H\left(y \over \rho(1-\rho) \right) 
\label{sca}
\end{equation}
where the function $H$ has the following properties
\begin{eqnarray}
H(y) \simeq  
  -(y-1)^2    \ \ \ \makebox{for } \ \ |y-1|   \ll 1  
\label{scaling2a}
\end{eqnarray}
\begin{eqnarray}
H(y) \simeq  
- (2 \sqrt{3} / (5 \sqrt{\pi}))  \ y^{5/2} 
  \ \ \  \makebox{for } \ \  y \to  + \infty  
\label{scaling2b}
\end{eqnarray}
\begin{eqnarray}
H(y) \simeq  
- (4 \sqrt{2 \pi} /  3 )  \ |y|^{3/2} \ \ \ \makebox{for } \ \  y \to -  \infty
\label{scaling2c}
\end{eqnarray}
Note that the difference between $\langle Y_t \rangle / t = p(N-p)/(N-1)$ and
$N\rho(1-\rho)$ used in (\ref{devia}) is responsible
for the fact that the maximum
of $f(y)$ is at $y=1$ as in (\ref{scaling2a}).
The scaling form (\ref{sca})  was obtained 
 (\ref{sol1a}-\ref{sol2a}) for $\alpha \sim N^{-3/2}$
and this means (\ref{devia}-\ref{scaling2c})
 that it is valid for $y$ of order 1.

For $y$ of order $N$, one expects from 
(\ref{largeap},\ref{largean}) 
the following  two scaling forms for $f(y)$:
\begin{equation}
f(y) \simeq  N H_+\left({y N^{-1}}\right) \ \ \ \makebox{for} \ \ \  y >0 
\label{hp}
\end{equation}
\begin{equation}
f(y) \simeq H_-\left({y N^{-1}}\right) \ \ \ \makebox{for} \ \ \  y <0 
\label{hm}
\end{equation}
As $y/(N\rho)$ is the deviation 
(from its most probable value $1-\rho$) in the velocity of the particles,
the extra factor $N$ in (\ref{hp}) has a simple interpretation:
to speed up the traffic, one needs to speed up all the particles  whereas
to slow down the traffic, it is sufficient to slow down a single particle. 
Note that if one considers that the scaling forms 
(\ref{sca},\ref{scaling2b},\ref{scaling2c})
 remain
valid for $1 \ll |y| \ll N$, it is possible to obtain
 from (\ref{scaling2b},\ref{scaling2c}) the small $z$ behavior
 of $H_+(z)$ and $H_-(z)$.

The distribution of $y$
is skew (\ref{cu3},\ref{mom3}) and  for large $y$ (but still $|y| \ll N$), 
it  decays  (\ref{scaling2b},\ref{scaling2c}) as the exponential of
a power law  with an exponent $5/2$ for
$y \to \infty$ and $3/2$ for $y \to -\infty$.
Skew distributions with similar tails have been observed in numerical
simulations  
\cite{HHZ,HH,bray,kmhh}
and partly understood by the replica approach
\cite{kardar,zhang,bo}. A precise comparison
with existing numerical data cannot  however be done
because in the present work, we consider a finite geometry (a ring
of $N$ sites, taking the long time limit first, and 
then the limit of an infinite lattice ($N\to \infty$)) whereas
the simulations had  been usually done for directed polymers on
an infinite lattice \cite{HHZ} (thus taking the limits in reverse order).
It would certainly be interesting to  try to extend the above Bethe ansatz
calculation to this case.
Other  extensions could be the calculation of
the diffusion constant \cite{dk},
the higher cumulants of $Y$ or the whole large
deviation function in more general cases such as
the partially asymmetric exclusion process  \cite{dk,Sandow,Kim,er}
or the case of several species 
\cite{DJLS}.

The similarity of  the expression (\ref{sol1a},\ref{sol2a})
of $G(\beta)$ as a function of $\beta$ 
with that of the pressure of a Fermi or a Bose gas in $3d$
was a surprise to us and it would be nice to have a more
direct explanation for it.
A Bose-Einstein condensation for the ASEP  has
already been found in the presence of quenched disorder \cite{evans}.
Here, the mechanism seems to be  different as instead of observing 
the Bose-Einstein condensation, one finds (\ref{sol1b},\ref{sol2b})
that $G(\beta)$ is analytically continued beyond the critical $\beta_0$.
\\ 
{\bf Acknowledgments:} 
We thank E. Speer for many very useful discussions.
  The research of JLL was supported by NSF Grant
95-23266 and AFOSR Grant 95-0159.


\begin{thebibliography}{10}
\bibitem{HHZ} T.~Halpin-Healy and Y.C.~Zhang,
{\it Phys. Rep.} {\bf 254}, 215 (1995)

\bibitem{bmp} J.P.~Bouchaud, M.~M\'ezard and G.~Parisi,
{\it Phys. Rev.} E {\bf 52}, 3656 (1995)

\bibitem{Lig} T.M.~Liggett, {\it  Interacting Particle Systems}
NY: Springer Verlag (1985)

\bibitem{Spo} H.~Spohn,  {\it Large Scale Dynamics of Interacting Particles}
NY: Springer Verlag (1991)

\bibitem{Krug} J.~Krug, 
{\it Phys. Rev. Lett.} {\bf 67}, 1882 (1991)

\bibitem{mrsb} P.~Meakin, P.~Ramanlal, L.M.~Sander and R.C.~Ball,
{\it Phys. Rev.} A {\bf 34}, 5091 (1986)

\bibitem{dhar} D.~Dhar, {\it Phase Transitions\/} {\bf 9}, 51 (1987)

\bibitem{gwa}
 L.H.~Gwa and H.~Spohn,
{\it Phys. Rev. Lett.}  {\bf 68}, 725 (1992); 
{\it Phys. Rev.} A {\bf 46}, 844 (1992)

\bibitem{DEHP}
 B.~Derrida and M.R.~Evans,
V.~Hakim and V.~Pasquier, {\it J. Phys.} A {\bf 26},
1493 (1993);
 B.~Derrida, M.R.~Evans,
 p 277  in {\it Non-equilibrium statistical mechanics
in one dimension},  ed V. Privman  (C.U.P.) 1997

\bibitem{dem} B.~Derrida, M.R.~Evans and D.~Mukamel, 
{\it J. Phys.} A {\bf 26}, 4911 (1993)

\bibitem{ss} R.B.~Stinchcombe  and  G.M.~Sch\"utz,
{\it Europhys. Lett.} {\bf 29}, 663 (1995) and
{\it Phys. Rev. Lett.} {\bf 75}, 140 (1995)

\bibitem{gs}   G.M.~Sch\"utz, 
 {\it J. Stat. Phys.} {\bf 88},  427  (1997)

\bibitem{janowsky} S.A.~Janowsky  and  J.L.~Lebowitz, 
{\it Phys. Rev.} A {\bf 45}, 618 (1992)

\bibitem{kt}
J.~Krug and L.H.~Tang, {\it Phys. Rev. } E {\bf 50}, 104 (1994)

\bibitem{kardar}
M.~Kardar, {\it Nucl. Phys.}  B {\bf 290} [FS20], 582 (1987)

\bibitem{zhang} Y.C.~Zhang, 
 {\it J. Stat. Phys.} {\bf 57}, 1123  (1989)

\bibitem{bo} J.P.~Bouchaud  and H.~Orland,  {\it J. Stat. Phys.} {\bf 61},
877 (1990)

\bibitem{HH}
T.~Halpin-Healy, {\it Phys. Rev.} A {\bf 44}, 3415 (1991)

\bibitem{bray} J.M.~Kim, A.J.~Bray and M.A.~Moore,
{\it Phys. Rev.} A {\bf 44}, 2345 (1991)

\bibitem{kmhh} J.~Krug, P.~Meakin and T.~Halpin-Healy, 
{\it Phys. Rev.} A {\bf45}, 638 (1992)

\bibitem{yc} V.~Yakhot and A. Chekhlov, 
{\it Phys. Rev. Lett.}  {\bf 77}, 3118 (1996)

\bibitem{huang}
 K.~Huang, {\it Statistical Mechanics}, John Wiley \& Sons 1987

\bibitem{dz}
B.~Derrida and  R.~Zeitak, {\it Phys. Rev. E}  {\bf 54}, 2513 (1996)

\bibitem{dk}
B.~Derrida and  K.~Mallick, {\it J.  Phys. } A {\bf 30}, 1031   (1997)

\bibitem{Sandow} S.~Sandow, {\it Phys. Rev.} E {\bf 50}, 2660 (1994)

\bibitem{Kim} D.~Kim, {\it Phys. Rev. } E {\bf 52}, 3512  (1995);    
 {\it J. Phys. } A {\bf 30}, 3817 (1997)

\bibitem{er}
F.H.L.~Essler and V.~Rittenberg, {\it J. Phys. } A {\bf  29}, 3375 (1996);
K.~Mallick and S.~Sandow, 
  {\it J. Phys. } A {\bf  30}, 4513 (1997)

\bibitem{DJLS} B.~Derrida, S.A.~Janowsky, J.L.~Lebowitz and E.R.~Speer,
{\it J. Stat. Phys.} {\bf 73}, 813 (1993)

\bibitem{evans} M.R.~Evans,
{\it Europhys. Lett.} {\bf 36}, 13 (1996);
  {\it J. Phys. } A {\bf  30}, 5669 (1997)

\end{thebibliography}
\end{document}